# Spatially Resolving the Enhancement Effect in Surface-Enhanced Coherent Anti-Stokes Raman Scattering by Plasmonic Doppler Gratings

*Lei Ouyang[†,‡,§], Tobias Meyer[†,‡], Kel-Meng See[⊥], Wei-Liang Chen[∥], Fan-Cheng Lin[⊥], Denis Akimov[†,‡], Sadaf Ehtesabi[‡], Martin Richter[‡], Michael Schmitt[†,‡], Yu-Ming Chang[∥], Stefanie Gräfe[‡], Jürgen Popp[†,‡], Jer-Shing Huang[†,‡,⊥,#,Δ,\*]*

[†] Leibniz Institute of Photonic Technology, Albert-Einstein Str. 9, 07745 Jena, Germany

[‡] Institute of Physical Chemistry and Abbe Center of Photonics, Friedrich-Schiller-Universität Jena, Helmholtzweg 4, D-07743 Jena, Germany

[§] School of Chemistry and Chemical Engineering, Huazhong University of Science and Technology, Wuhan 430074, China

[⊥] Department of Chemistry, National Tsing Hua University, 101 Sec. 2, Kuang-Fu Road, Hsinchu 30013, Taiwan

[∥] Center for Condensed Matter Sciences, National Taiwan University, Taipei 10617, Taiwan

[#] Research Center for Applied Sciences, Academia Sinica, 128 Sec. 2, Academia Road, Nankang District, Taipei 11529, Taiwan

[Δ] Department of Electrophysics, National Chiao Tung University, Hsinchu 30010, Taiwan

\*Corresponding Author, E-mail: jer-shing.huang@leibniz-ipht.de

**Abstract**

Well-designed plasmonic nanostructures can mediate far and near optical fields and thereby enhance light-matter interactions. To get the best overall enhancement, structural parameters need to be carefully tuned to obtain the largest enhancement at the input and output frequencies. This is, however, challenging for nonlinear light-matter interactions involving multiple frequencies because obtaining the full picture of structure-dependent enhancement at individual frequencies is not easy. In this work, we introduce the platform of plasmonic Doppler grating (PDG) to experimentally investigate the enhancement effect of plasmonic gratings in the input and output beams of nonlinear surface-enhanced coherent anti-Stokes Raman scattering (SECARS). PDGs are designable azimuthally chirped gratings that provide broadband and spatially dispersed plasmonic enhancement. Therefore, they offer the opportunity to observe and compare the overall enhancement from different combinations of enhancement in individual input and output beams. We first confirm PDG's capability of spatially separating the input and output enhancement in linear surface-enhanced fluorescence and Raman scattering. We then investigate spatially resolved enhancement in nonlinear SECARS, where coherent interaction of the pump, Stokes, and anti-Stokes beams is enhanced by the plasmonic gratings. By mapping the SECARS signal and analyzing the azimuthal angle-dependent intensity, we characterize the enhancement at individual frequencies. Together with theoretical analysis, we show that while simultaneous enhancement in the input and output beams is important for SECARS, the enhancement in the pump and anti-Stokes beams plays a more critical role in the overall enhancement than that in the Stokes beam. This work provides an insight into the enhancement mechanism of plasmon-enhanced spectroscopy, which is important for the design and optimization of plasmonic gratings. The PDG platform may also be applied to study enhancement mechanisms in other nonlinear light-matter interactions or the impact of plasmonic gratings on the fluorescence lifetime.

Plasmonic nanostructures can function as optical nanoantennas to mediate optical near- and far-field and thereby enhance specific light-matter interactions.[1] Applications of well-engineered nanostructures include plasmon-enhanced catalysis,[2] plasmonic photovoltaic devices,[3] plasmonic sensors,[4,5] and plasmon-enhanced spectroscopy.[6-11] In plasmon-enhanced spectroscopic methods, the overall optical signal enhancement is in principle proportional to the enhancement in the input (excitation) and output (emission or scattering).[12] In practice, a single enhancement factor determined by the localized surface plasmon resonance (LSPR) can be used to estimate the overall enhancement if the LSPR is significantly broad to well cover the input and output frequencies. In this case, the overall enhancement is proportional to the fourth power of the electric field enhancement at the frequency of the LSPR.[13] This approximation is, however, insufficient for light-matter interactions involving multiple frequencies with a large spectral difference or when the nanostructures exhibit complex resonance spectra. A complete understanding of the plasmonic enhancement of the structure in individual beams with respect to critical structural parameters is important for the design of the best nanostructures. To understand the roles of enhancements at individual input or output frequencies, it is necessary to disentangle the enhancement effect so that their contributions to the overall enhancement can be separately quantified. This is particularly important for plasmon-enhanced nonlinear light-matter interactions involving coherent interaction of multiple beams at different frequencies. Coherent anti-Stokes Raman scattering (CARS) is one such nonlinear light-matter interaction.[14-18] In the simplest case of two-color CARS, coherent interaction of two input beams at the frequencies of the pump and the Stokes results in one output beam at the anti-Stokes frequency. In a classical model, the enhancement in the CARS signal compared to spontaneous Raman scattering is due to the coherent enhancement, *i.e.*, the coherent population of molecules to a specific vibrational state defined by the energy difference between the pump and Stokes beams.[18] In addition to this coherent enhancement effect, the CARS signal can be further increased by chemical enhancement of the substrate[17] and surface enhancement effect from random or well-designed nanostructures.[10, 14-16, 19-31] In this work, we focus on the surface enhancement effect of the plasmonic nanostructures in surface-enhanced coherent anti-Stokes Raman scattering (SECARS). The coherent enhancement and chemical enhancement involved in the CARS process are not directly relevant to the main observation and thus not discussed in this work. Figure 1a schematically summarizes the transitions involved in the fluorescence, Raman scattering, and CARS.

In principle, a suitable plasmonic nanostructure for SECARS should serve two purposes. First, it should serve as an efficient receiving nanoantenna to enhance the input beams, *i.e.*, to concentrate the incoming pump and Stokes beams into enhanced optical near fields localized at the position of the molecules. Secondly, it should serve as a transmitting optical antenna to enhance the output, *i.e.*, to provide efficient radiative channels at the anti-Stokes frequency for the radiative decay of the local energy. The detected SECARS signal is thus proportional to the field enhancement at the pump ($E_p$), Stokes ($E_S$), and anti-Stokes ($E_{aS}$) frequencies, *i.e.*, $I_{SECARS} \propto |E_p|^4 \times |E_S|^2 \times |E_{aS}|^2$.[15] Although this formula is commonly used to account for the surface enhancement seen in SECARS, it does not reflect the coherent nature of CARS. From a quantum mechanical point of view, the CARS process can be divided into two parts. In the first part, molecules in the ground state are efficiently populated to the excited vibrational state by coherent interaction of the pump and Stokes photons. In the second part, pump photons are scattered by the coherently vibrating molecules into photons at the anti-Stokes frequency. According to the pictures based on antenna theory and quantum mechanics, we can outline three important points for the design of the nanostructures to provide the highest surface enhancement in SECARS. First, since CARS involves coherent interaction of the pump and Stokes photons, it is important for the nanostructures to simultaneously provide near-field intensity enhancement for the pump and Stokes beams. Secondly, surface enhancement from the structure in the pump has a greater effect than that in the Stokes since the pump photons are involved in both parts of the CARS process. This is also the reason for the quadratic dependence of the CARS signal on the pump power. Thirdly, the structure's enhancement effect in the emission at the anti-Stokes frequency plays a decisive role in the final CARS signal because it determines how efficient the local energy on the polarized molecules can decay radiatively into the inelastically scattered anti-Stokes photons.

To verify these principles and design plasmonic nanostructures with the best near- and far-field properties for SECARS, one needs to investigate the surface enhancement effect at individual frequencies with respect to critical geometrical parameters of the nanostructures. Although a wide diversity of plasmonic nanostructures has been employed to provide enhancement, including rough metallic films,[19] colloidal nanoparticles,[20-22, 24] and rationally designed complex nanostructures,[14, 25, 27, 30] most of the works only presented the overall enhancement without systematic experimental characterization of the enhancement at individual frequencies with respect to structural parameters. A clear guideline for nanostructure design for SECARS is still missing. Several published works on SECARS even showed different design principles. For example, Steuwe et al.[25] demonstrated CARS signal enhancement by a factor of $10^5$ by using a nanostructured gold film with three resonant peaks matching the pump, Stokes, and anti-Stokes frequencies. Similarly, He et al.[28] presented a plasmonic nanoassembly consisting of three asymmetric gold disks to maximize the enhancement. They also theoretically investigated a crisscross dimer array for SECARS. The claim is that nanostructures with scattering peaks matching the three frequencies of CARS would lead to maximal enhancement.[29, 31] Differently, Zhang *et al.*[15] used plasmonic nanoquadrumers to improve the sensitivity of CARS down to the single-molecule level. Their nanostructure was designed to exhibit Fano-like resonance due to the interference of narrow sub-radiant mode and broad super-radiant mode. The Fano plasmonic resonator shows two scattering maxima at the Stokes and the anti-Stokes frequencies and a dip at the pump frequency, instead of a peak like in He's works. Although these works all show SECARS, the design principles are different and the enhancement effect in individual beams was not experimentally characterized against critical structural parameters. A complete understanding of the role of the structure in the enhancement at the pump, Stokes, and anti-Stokes frequencies is necessary for the optimization of the structure.

In this work, we employed plasmonic Doppler gratings (PDGs),[5, 32] azimuthally chirped plasmonic gratings, to systematically study the effect of different combinations of surface enhancement at different frequencies. PDGs allow us to spatially disentangle the enhancement at different frequencies by observing the angle distribution of the SECARS signal intensity. With spectral mapping, we spatially resolve the input and output enhancement in linear surface-enhanced spectroscopy (surface-enhanced fluorescence, SEF, and surface-enhanced Raman scattering, SERS). We show that PDG is a perfect platform to spatially separate different combinations of surface enhancement effects, namely no enhancement, enhancement in either the input or output, or enhancement in both. Further, we apply PDG to study the surface enhancement in the more complex nonlinear SECARS. We map the intensity of SECARS on a PDG and analyze the intensity with respect to azimuthal angles. This allows us to attribute the observed SECARS signals to specific combinations of enhancement in the pump, Stokes, or anti-Stokes beams.

**Results and discussion**

**Working principle of PDG.** Metallic gratings are effective yet simple nanostructures to couple free-space photons into surface plasmons. Since metallic gratings offer both the effects of localized surface plasmon resonance and grating resonance, they have been used as an ideal platform for enhancing nonlinear light-matter interactions.[23, 33-37] The resonance frequency of a plasmonic grating can be easily tuned by changing its periodicity and the incident angle of illumination. Compared to a flat gold surface, a near-field intensity enhancement of more than three orders of magnitude has been demonstrated on a gold grating for four-wave mixing.[23] Plasmonic gratings consisting of concentric rings have also been used to demonstrate the near-field enhancement effect in SERS.[38] However, most reported plasmonic gratings only have one single periodicity, making it difficult to simultaneously provide enhancement at multiple frequencies. To address this issue, we employed PDGs, which are azimuthally chirped gratings consisting of a series of non-concentric circular slits or grooves on a metallic film, to provide broadband and azimuthally chirped grating periodicities.[5, 32, 39] The rings of a PDG mimic the wavefronts of a moving point source that exhibits the Doppler effect. A PDG thus provides continuously chirped grating periodicities along the in-plane azimuthal angle (Figure 1b). The trajectory of the n$^{th}$ ring of a PDG is

$$(x - nd)^2 + y^2 = (n\,\Delta r)^2, \qquad (1)$$

where $\Delta r$ is the radius increment, $d$ is the center displacement of a circular ring relative to its adjacent rings, and $n$ is the serial number of the rings. Here, we focus on the cases with $d < \Delta r$. The PDG thus possesses a chirped periodicity that depends on the in-plane azimuthal angle ($\varphi$),

$$P_{\Delta r,d}(\varphi) = \left| d\cos\varphi \pm \sqrt{(d^2 \cos 2\varphi + 2\Delta r^2 - d^2)/2} \right|, \qquad (2)$$

The periodicity of the gratings in a PDG varies continuously from $\Delta r+d$ to $\Delta r-d$ as the in-plane azimuthal angle increases from 0° to 180°. The range of periodicity can be freely designed by choosing the suitable $\Delta r$ and $d$. Taking this azimuthal angle-dependent periodicity into the photon-plasmon phase-matching condition, an azimuthal angle-dependent resonance wavelength is obtained, [5, 32]

$$\lambda_0 = \frac{\left|d\cos\varphi \pm \sqrt{(d^2\cos 2\varphi + 2\Delta r^2 - d^2)/2}\right|}{m} \left(\sqrt{\frac{\varepsilon_m \cdot n_d^2}{\varepsilon_m + n_d^2}} - n_d \sin\alpha\right), \qquad (3)$$

where $m$ is the resonance order, $\alpha$ the incident angle, $\varepsilon_m$ the permittivity of the metal, and $n_d$ the index of the dielectric surrounding. Eq. (3) allows for analytically predicting the range of in-plane azimuthal angle for efficient far-to-near field coupling at a specific frequency and incident/emission angle. If a finite angle range is allowed for the incident/emission angle ($\alpha$) due to, for example, the finite numerical aperture (NA) of a microscope objective, the corresponding azimuthal angle for photo-to-plasmon coupling would expand into a sector area. Depending on the design of PDGs, the sector areas for different frequencies can overlap, offering the possibility to simultaneously provide enhancement for multiple beams at different frequencies. By imaging the spatial distribution of the SEF/SERS and SECARS signal and analyzing the enhancement as a function of the azimuthal angle range, it is possible to spatially separate the enhancement effect at different frequencies.

**Design and fabrication of PDG.** In this work, we designed and fabricated two PDGs to demonstrate the spatial disentanglement of the enhancement effects at the input and output frequencies for linear SEF/SERS and nonlinear SECARS. For the SEF/SERS, the excitation wavelength is 633 nm and the molecule is Rhodamine 6G (R6G), which exhibits a broad Raman response from 100 cm$^{-1}$ to 3000 cm$^{-1}$, corresponding to Stokes Raman scattering in the spectral range between 640 nm and 770 nm. Accordingly, we have chosen Δr = 450 nm and d = 140 nm for the PDG in order to observe all possible combinations of enhancement effects on one single PDG. For SECARS, we have chosen 4-Aminothiophenol (4-ATP) as the sample and target to its characteristic peak at 1070 cm$^{-1}$. The reasons for choosing 4-ATP as the sample molecule are as follows. First, it is a widely used SECARS probe, which has a relatively simple Raman spectrum with strong CARS peaks.[15] Secondly, 4-ATP molecules bind strongly to the

Au surface through the Au-S bond. This allows the formation of a molecular monolayer[40] which guarantees the homogeneity of the molecules. Thirdly, according to our calculations based on density functional theory (DFT), the binding of the 4-ATP molecules to the gold surface can greatly enhance the characteristic Raman peak at 1070 cm$^{-1}$, which is the targeted peak in our SECARS experiment (see Supporting Information). Our laser system provides the pump and Stokes beams at 955 nm and 1064 nm, respectively. Therefore, the anti-Stokes peak at 1070 cm$^{-1}$ of 4-ATP is expected to show up at 867 nm. To cover the wide frequency range from the anti-Stokes beam at 867 nm to the Stokes beam at 1064 nm with one PDG, we have chosen Δr = 700 nm and d = 400 nm for the PDG design to make sure that all the frequencies of interest are covered. The SEM images of the obtained PDGs are shown in Figures 1c and 1d. The reflection images of the PDG structures under white light illumination show characteristic rainbow color distributions due to the azimuthally chirped periodicity (insets of Figures 1c and 1d). The color distributions are completely different because these two PDGs are optimally designed for different spectral ranges for SEF/SERS and SECARS.

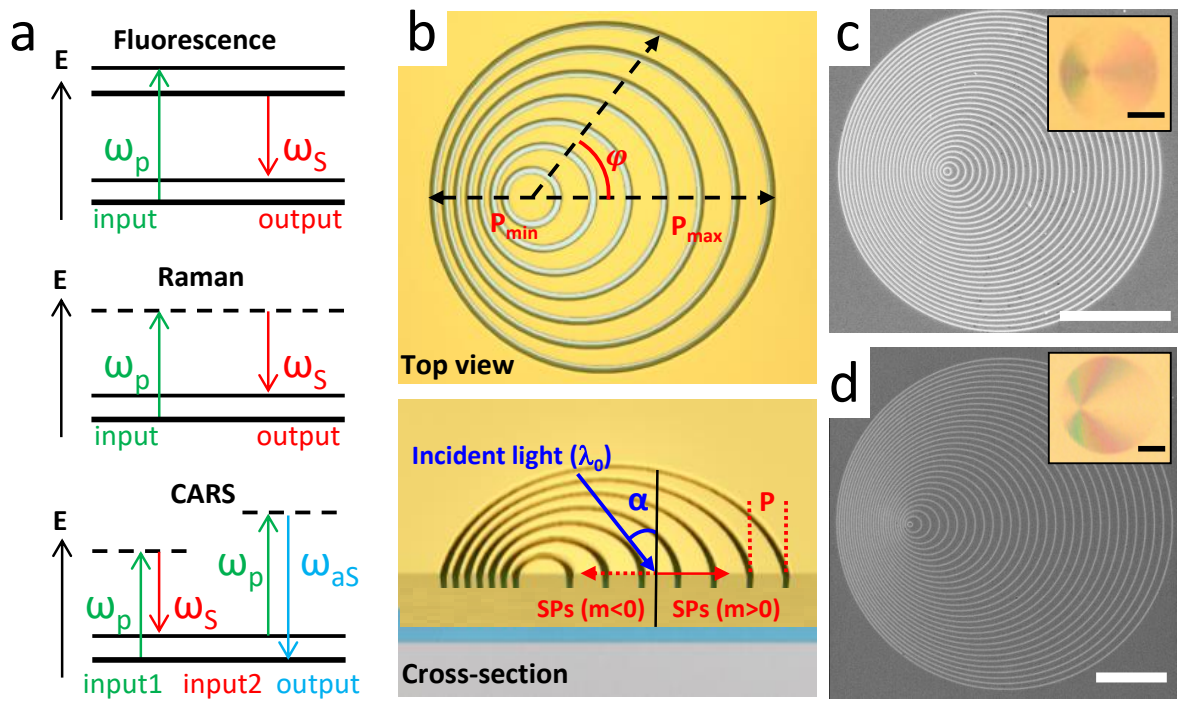

Figure 1 (a) Schematic illustration of the transitions in the processes of photon-excited fluorescence, Stokes Raman scattering, and CARS. (b) The design of PDG shown in the top view (upper panel) and the tilted view (lower panel). φ is the in-plane azimuthal and α is the incident angle relative to the surface normal. (c, d) SEM images of the PDG structures used in this work for (c) SEF/SERS with $\Delta r$ = 450 nm, $d$ = 140 nm and for (d) SECARS with $\Delta r$ = 700 nm, $d$ = 400 nm. Their optical reflection images are shown in the insets. All scale bars are 10 μm.

**Spatially resolved enhancement on PDG structure.** The surface enhancement in surface-enhanced spectroscopy can be understood as a result of a two-step process. In the first step, the plasmonic nanostructure acts as a receiving optical antenna, which localizes the far-field light at the input frequency to a highly concentrated optical near field. The intense optical near field enhances the vibrational polarizability of the molecules within the hot spot by one to three orders of magnitude, mainly due to the strong mutual excitation between the induced dipole of molecules and the dipole (and even multipoles) of the nanostructures.[41] In the second step, the plasmonic structure serves as a transmitting optical nanoantenna to provide radiative channels for the near-field optical energy on the molecule to decay into far-field electromagnetic waves at the fluorescence or Raman scattering frequency. The overall surface enhancement is proportional to the square of the local electric field enhancement at the frequencies of the input and output.[41] For linear surface-enhanced processes such as SEF and SERS, the surface enhancement ($E_{Surface}$) relative to spontaneous emission and spontaneous Raman scattering is the product of the square of the local field enhancement at the excitation frequency ($E_{ex}$) and emission frequency with a Stokes shift ($E_S$),

$$E_{Surface} = |E_{ex}|^2 \times |E_S|^2 \qquad (4)$$

For the case of nonlinear SECARS, the surface enhancement factor can be defined as $I_{SECARS}/I_{CARS}$, which can be obtained by comparing the signal of SECARS with that of

CARS.[10] The surface enhancement factor $E_{SECARS/CARS}$ is proportional to the fourth power of the local field enhancement at the pump frequency ($E_p$), the square of that at the Stokes frequency ($E_S$) and anti-Stokes frequency ($E_{aS}$),

$$E_{SECARS/CARS} = |E_p|^4 \times |E_S|^2 \times |E_{aS}|^2 \qquad (5)$$

To spatially disentangle the contribution from enhancement at different frequencies, we recorded and analyzed the azimuthal angle-resolved signal enhancement on the rationally designed PDGs. According to Eq. (3), the resonance order $m$ and the incident angle $\alpha$ also contribute to the determination of the azimuthal angles for photon-to-plasmon coupling at a specific frequency. Since typical objectives without any beam stop at the back focal plane provide a full range of incident angle from 0° (normal incidence) to $\alpha$ (limited by the numerical aperture of the objective), a PDG naturally provides a finite range of azimuthal angle (sector areas) for photon-to-plasmon coupling at a specific frequency. Figure 2 illustrates such sector areas on the PDG for SEF/SERS (two frequencies are involved), and for SECARS (three frequencies are involved). The overlapping regions between the excitation enhancement sectors and the emission enhancement sectors are expected to give the highest signal. For SEF/SERS, there are only three cases of enhancement combinations, namely enhancement in both the input and output, enhancement in either one of the two beams, or no enhancement for both, as schematically illustrated in Figure 2a. For SECARS, the situation is more complex. As shown in Figure 2b, the PDG is designed to provide an enhancement in the pump, Stokes, and anti-Stokes frequencies. The corresponding sector areas for enhancement at these three frequencies partially overlap, resulting in spatially separated areas for various combinations of the enhancement. The strongest SECARS intensity is expected to be observed in the areas where the gratings promote photon-to-plasmon coupling at all three frequencies (the violet sectors in Figure 2b). For the areas where pump and Stokes are enhanced but anti-Stokes is not, moderate SECARS enhancement is expected. However, in the areas where only one of the pump and Stokes beams is enhanced, the enhancement of the CARS signal is expected to be very low because CARS requires coherent interaction of the pump and the Stokes photons at the position of the molecules (see Figure 1a). Missing the enhancement in any one of the two input beams would lead to very weak overall enhancement because the preparation of the coherently vibrating molecules for enhanced scattering becomes inefficient. As for the enhancement at the anti-Stokes frequency, it is independent of the enhancement condition of the two input beams. The nanostructure simply serves as an emitting antenna that provides additional radiative decay channels and thereby increases the emission efficiency at the anti-Stokes frequency.

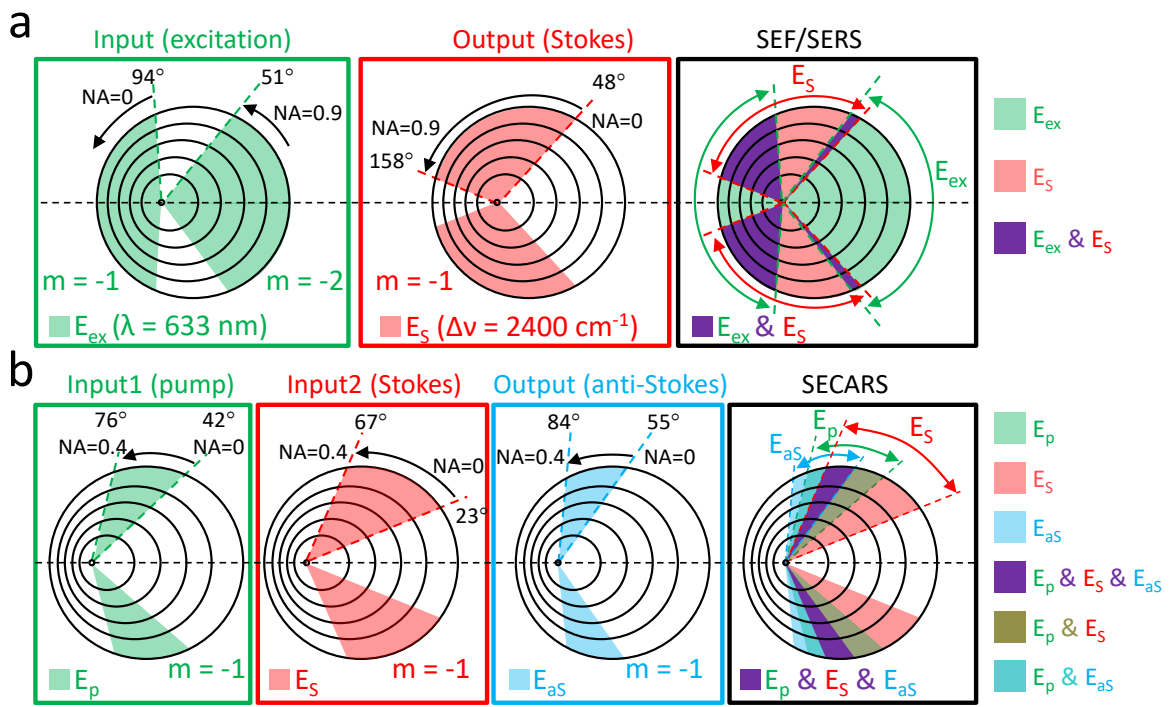

Figure 2. Schematic illustration of the sector areas on a PDG for (a) SEF and SERS (PDG $\Delta r$ = 450 nm, $d$ = 140 nm; Stokes shift = 2400 cm$^{-1}$; $\lambda_{ex}$ = 633 nm) and (b) SECARS (PDG $\Delta r$ = 700 nm, $d$ = 400 nm; Anti-Stokes shift = 1070 cm$^{-1}$; $\lambda_{pump}$ = 955 nm; $\lambda_{Stokes}$ = 1064 nm; $\lambda_{anti-Stokes}$ = 867 nm). The range of azimuthal angle for each sector area is analytically calculated. The expected situations of enhancement are summarized on the right side of Figure 2.

**SEF/SERS enhancement on PDGs.** To demonstrate spatially resolved enhancement effect in the input and output of linear surface-enhanced spectroscopy, we studied the SEF/SERS from R6G molecules in a PMMA film coated on an optimally designed PDG. The thickness of the film is about 10 nm.[42] This avoids the quenching of the fluorescence because the majority of the R6G molecules are not in direct contact with the gold surface. The focal spot of a circularly polarized CW laser at 633 nm (averaged power: 0.5 mW) was scanned across the area of PDG and unpatterned flat gold surface. For the enhancement at the input frequency, we first examined the reflection of the excitation laser by the PDG (Figure 3a). Compared to the analytically calculated coupling angles for 633 nm excitation focused by an air objective with a NA of 0.9 (Figure 3b), the experimental reflection image shows dark bands at the in-coupling azimuthal angles predicted by the analytical model due to effective conversion of the illumination into surface plasmons. Differences between the experimental image and analytical prediction are found in azimuthal angle range between $\varphi = \pm 35°$. Within this range, the grooves of the grating are clearly resolved because the periodicities ($\geqslant 557$ nm) are larger than the size of the laser focal spot (~430 nm). This means that the incoming light sees individual grooves instead of the grating. As a result, the coupling efficiency is low and the reflection intensity restores. The grating within this azimuthal angle range is, therefore, not suitable for spatially resolving the enhancement effect. Apart from this, the PDG converts the excitation into an enhanced optical near field at the expected angles. In the following, we ignore the signals from gratings within this range. Next, we performed confocal spectral mapping by recording the emission spectrum at each excitation position. This allows us to obtain the overall emission intensity image as well as the spectrally resolved emission images at wavelengths corresponding to specific Stokes shifts. The emission spectra taken from three representative locations on the sample are shown in Figure 3d. In the bright region (black arrow, $\varphi = 180°$, Figure 3c), clear Raman peaks are observed on a broad continuous fluorescence background due to strong signal enhancement by the plasmonic gratings. At the other two positions, *i.e.*, the area with grating periodicity larger than the focal spot size (green arrow, φ = 0°, in Figure 3c,) and the unpatterned area (red arrow in Figure 3c), the emission intensity is rather weak and the Raman peaks are difficult to observe. Figure 3e shows the spectrally resolved SEF/SERS images at various Stokes shifts between 400 and 2400 cm$^{-1}$. The upper panels of Figure 3e show the analytically predicted sector areas of gratings which provide the enhancement in the excitation and emission at the Stokes shift of 400 to 2400 wavenumbers. The spatially localized SERS enhancement on PDG can be classified into four regions: (i) excitation-enhancement regions ($E_{ex}$), (ii) emission-enhancement region at specific Stokes frequencies ($E_S$), (iii) simultaneous excitation- and emission-enhancement regions ($E_{ex}$ and $E_{Stokes}$) and (iv) no enhancement region. With the knowledge of the PDG design ($\Delta r$ = 450 nm, $d$ = 140 nm), the excitation wavelength (633 nm), and the numerical aperture of the objective (NA = 0.90), these four regions can be analytically predicted. The strongest fluorescence and SERS signal are expected to show up at the sector regions where both the input and the output beams are enhanced. It is clear that at the grating area with the smallest periodicity (φ = 180°) the PDG enhances the signals up to the Stokes shift of 2400 cm$^{-1}$. As the Stokes shift increases, the grating suitable for enhancing the emission shifts to larger periodicity, *i.e.*, φ shifts toward 0. Starting from 2000 cm$^{-1}$, additional bright bands emerge around φ = ± 49° (see yellow arrows in Figure 3e) due to the overlap of the enhancement sector areas for emission ($E_S$, m = -1) and excitation ($E_{ex}$, m = -2). A movie of the SEF/SERS images as a function of the Stokes shift is provided in the Supporting Information. The experimental emission intensity distributions (lower panel of Figure 3e) agree very well with the analytically predicted overlapping sector areas (violet areas in the upper panels of Figure 3e). Difference between the experimental and analytically predicted intensity angle distributions between φ = ± 35° is because the gratings within this angle range have periods larger than the focal spot size. Therefore, the excitation sees individual grooves instead of the gratings, leading to low enhancement in the excitation and low overall signal. Nevertheless, the PDG demonstrates spatially resolved enhancement effect and confirms the importance of having enhancement simultaneously in the input and output for SEF and SERS. Note that the analytically calculated sector areas in the upper panels of Figure 3(e) are calculated based on the photon-to-plasmon momentum matching equation and thus only indicate the best coupling angles. It does not take into account any broadening effect that can lead to the finite bandwidth of the plasmon resonance. Therefore, one should keep in mind that the sharp boundaries of the sectors indeed have a finite width due to the bandwidth of the plasmon resonance.[32] As a result, around 2000 cm$^{-1}$, the enhancement already emerges in the experimental before the calculated boundaries overlap.

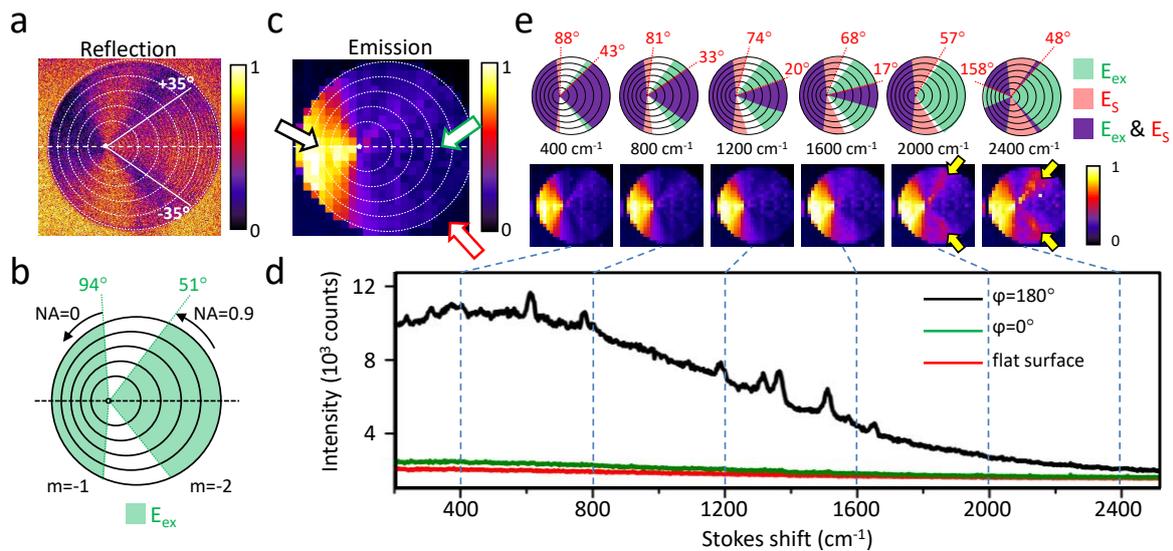

Figure 3 (a) Image of reflection of the excitation at 633 nm by the PDG. The dashed line is the horizontal axis of the PDG. The white lines mark the boundaries of the azimuthal angles range between φ = ± 35°. Within this range, the grating periodicity is larger than the size of the excitation focal spot, and the grooves of PDG are clearly resolved. (b) Analytically predicted coupling sector areas of the 633-nm excitation *via* grating resonance orders m = -1 and m = -2. Black arrows point toward the boundaries of the sector areas due to the numerical aperture of the air objective (NA = 0.9). (c) Confocal intensity map (full-spectrum) of the emission of R6G on the PDG. The black, green, and red hollow arrows indicate the positions where the spectra shown in (d) were obtained. (d) Emission spectra recorded at three different locations on the PDG corresponding to strongly enhanced ($\varphi = 180°$, black spectrum, black hollow arrow in (c)) and weakly enhanced grating areas ($\varphi = 0°$, green spectrum, green hollow arrow in (c)), as well as the unpatterned flat gold surface (red spectrum, red hollow arrow in (c)). (e) Upper panels: analytically predicted enhancement sector areas for the excitation at 633 nm (green), emission at various Stokes shifts from 400 to 2400 cm$^{-1}$ (red), and the overlapping areas of the excitation and emission enhancement (violet). Lower panels: experimental emission intensity maps at various Stokes shifts. The averaged excitation power of the 633-nm CW laser is 0.5 mW.

**SECARS enhancement on PDG**. Having demonstrated PDG's capability to spatially resolve the enhancement effect in the input and output, we further study more complex nonlinear SECARS, where four photons at three different frequencies are involved. Ideally, the grating should serve as a receiving antenna for the two input beams and a transmitting antenna for the output beam. The strongest SECARS signal is thus expected to be observed at the angles where the gratings enhance the local field at the pump and Stokes frequencies and promote the emission at the anti-Stokes frequency. The analytically predicted resonant sector areas for the pump (955 nm), Stokes (1064 nm), and anti-Stokes beam (867 nm) are shown in Figure 2b. The sector areas responsible for enhancing the three beams overlap between $\varphi = 55°$ and $\varphi = 67°$, as indicated by the violet sector in Figure 2b. Therefore, the largest SECARS signal is expected to be obtained in this angle range. The experimental intensity map of SECARS of the 4-ATP molecular monolayer on the PDG is shown in Figure 4a. A clear azimuthal angle-dependent SECARS intensity distribution is observed between $\varphi = 55°$ and $\varphi = 84°$. The SECARS signal measured in this sector area shows a clear quadratic dependence on the pump power and linear dependence on the Stokes power (Figure 4b), confirming the nonlinearity of the CARS signal. To compare with the analytically predicted angle distributions shown in

Figure 2b, the boundaries of the sector areas responsible for the enhancement in the pump, Stokes, and anti-Stokes beams are marked in Figure 4a. The overlap of these three sectors results in various combinations of enhancement effect, as indicated in Figure 4a. Clearly, a strong SECARS signal is observed in the sector area between $\varphi = 55°$ and $\varphi = 67°$, where the gratings are expected to enhance all three beams in CARS. A close look into the experimental data reveals that the spatial distribution of the SECARS signal extends counterclockwise into the neighboring sector between $\varphi = 67°$ and $\varphi = 76°$, where pump and anti-Stokes are enhanced, and vanishes at about $\varphi = 84°$, which is the boundary of the sector for anti-Stokes enhancement. Interestingly, in the clockwise direction, the SECARS signal drops quickly, even though the neighboring sector area between $\varphi = 42°$ and $\varphi = 55°$ is expected to enhance both the pump and the Stokes beams. In fact, the SECARS signal is only observed between $\varphi = 55°$ and $\varphi = 84°$, coinciding with the sector area for enhancement in the anti-Stokes output. Comparing the SECARS signal from the sector area between 67° and 76° (for pump and anti-Stokes enhancement) with that from the sector area between 42° and 55° (for pump and Stokes enhancement), it becomes obvious that enhancement in the anti-Stokes beams plays a decisive role in the overall SECARS signal. Another interesting feature to know is that the SECARS signal from the sector area between $\varphi = 67°$ and $\varphi = 76°$ (for pump and anti-Stokes enhancement) is almost the same as that in the sector between $\varphi = 55°$ and $\varphi = 67°$ (for enhancement in all three beams), suggesting that the enhancement in the Stokes beam is less significant.

To understand these features, we further perform quantitative analysis on the enhancement effect. Since the photon-to-plasmon coupling angles are calculated based on the photo-to-plasmon phase-matching condition, it does not provide quantitative information on the enhancement factor. To obtain the enhancement factor, one also needs to take into account the incident/emitting angle and the morphology of the grating grooves. Therefore, FDTD simulations were performed to simulate the near-field enhancement for the input beams at various incident angles and emission enhancement for the output beam at different emitting angles. The range of the incident and emitting angle is determined by the numerical aperture of the objective. For the enhancement in the pump and Stokes beam, near-field intensity enhancement is simulated under the illumination of plane waves at all allowed incident angles. The enhancement factor is then calculated by normalizing it to the values obtained on a flat gold surface under the same illumination condition. For the enhancement in the output at the anti-Stokes frequency, a dipole source with an in-plane dipole orientation perpendicular to the grooves is placed in the close vicinity of the grating, and the emission power within the collectible emission angle is simulated. The enhancement factor in the anti-Stokes output is then compared with a single dipole on the flat gold surface. Details of the FDTD simulations on the enhancement in SECARS can be found in the Supporting Information.

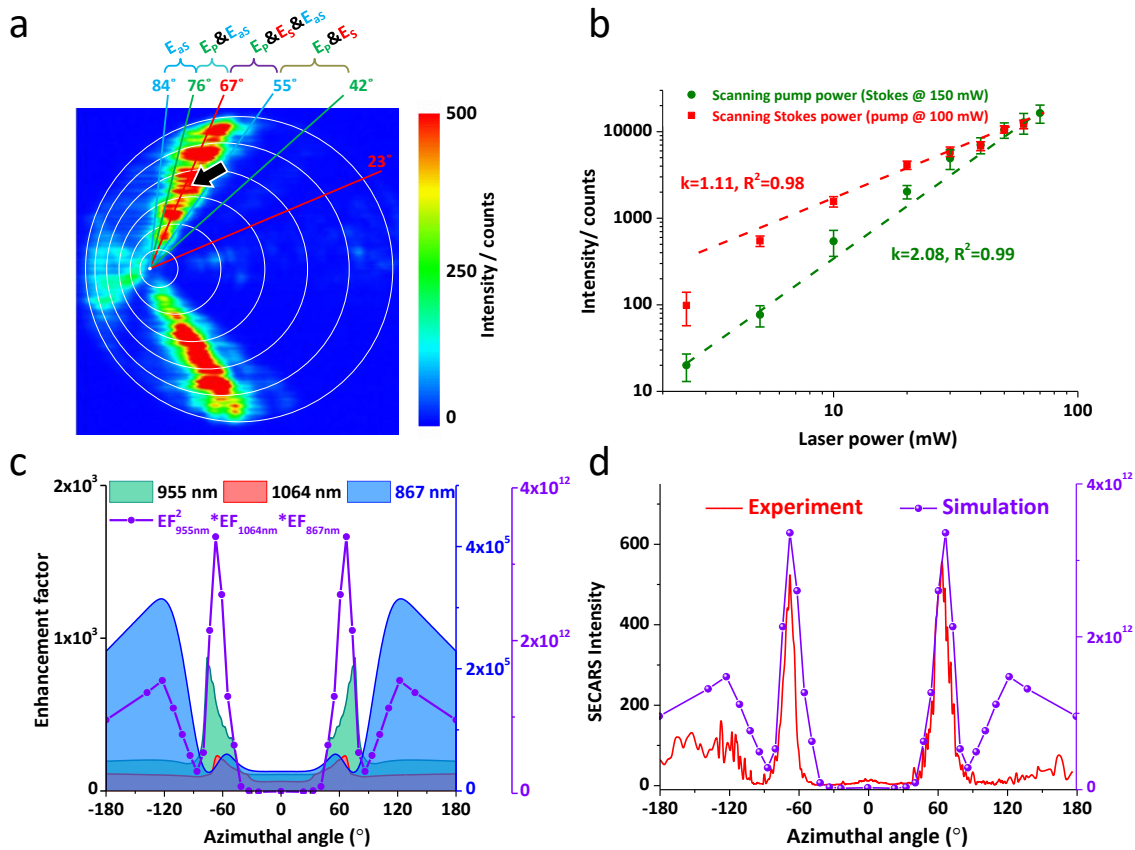

Figure 4. (a) Experimental SECARS intensity map recorded at the anti-Stokes wavelength (867 nm), *i.e.*, 1070 cm$^{-1}$ blue-shifted from the pump at 955 nm. (b) Maps of exponent $k$ for the power dependence on the pump (left) and Stokes (right). (c) Simulated enhancement factor at the pump (green, scale to the left) and Stokes frequencies (red, scale to the left), as well as the enhancement factor in the scattering at the anti-Stokes frequency (light blue, scale to the right). Overall enhancement calculated using $I_{SECARS} \propto E_p^4 * E_S^2 * E_{aS}^2$ is also plotted with a separate scale on the right side. (d) Experimental SECARS intensity (red, scale to the left) and simulated SECARS enhancement factor (violet, scale to the right) as functions of the azimuthal angle.

Figure 4c shows the simulated input enhancement for the pump and Stokes beams, and the output enhancement at the anti-Stokes frequency. Note that the scale for the enhancement in the anti-Stokes beam is on the right side of Figure 4c and is about three orders of magnitude larger than that for the pump and Stokes beams. The overall field enhancement calculated using $I_{SECARS} \propto |E_p|^4 * |E_S|^2 * |E_{aS}|^2$ is also plotted in Figure 4c. The simulations show that the enhancement in the anti-Stokes has a major contribution to the overall enhancement, whereas the enhancement in the Stokes beam has a relatively small impact. This explains the experimental observation that enhancement in the Stokes beam is less significant. The experimental SECARS intensity distribution is further transformed into the azimuthal distribution profile (Supporting Information) and plotted with the simulated overall enhancement in Figure 4d. The experimental and simulated angle distribution profiles agree well in the azimuthal angle and the shape of the intensity angle profile. According to the simulated enhancement profiles in Figure 4c, the enhanced signal around azimuthal angle φ = ±67° is mainly due to the enhancement in the pump and anti-Stokes beams and the SECARS signal seen between φ = ±120° and ±180° is majorly due to the emission enhancement at the anti-Stokes frequency. This confirms the important role in the pump beam due to the quadratic power dependence of the SECARS on the pump. It also shows the critical role of the nanostructure as an emitting antenna which provides the enhancement in the out-going anti-Stokes beam. The analysis shown here provides information on the enhancement contribution from individual input and output beams and allows us to gain insight into the enhancing mechanism beyond the overall enhancement.

**Conclusions**

We have demonstrated the capability of PDGs as a designable platform to spatially separate the enhancement in the input and output in linear (SEF/SERS) and nonlinear surface-enhanced spectroscopy (SECARS). Due to the azimuthally chirped grating, various combinations of enhancement in input and output beams can be found at different angle ranges of a single PDG, offering the possibility to study the enhancement mechanism concerning the structural parameters. We demonstrated the capability of PDG to spatially resolve enhancement in the input and output beams in SEF/SERS and SECARS. For SEF/SERS, the highest signal is obtained from the grating areas where both the excitation and the emission are enhanced. For SECARS, we found that the enhancement in the pump beam and the anti-Stokes beam is decisively important, whereas the enhancement in the Stokes beam plays a less important role. PDG is an effective platform to study the impact of the enhancement in individual input and output beams. A further study on the plasmonic enhancement effect would be to introduce the "time" dimension. For example, fluorescence lifetime imaging can be performed on emitter coated PDGs. Since the enhancement in the excitation, emission, and non-radiative quenching would impact the fluorescence lifetime differently, it is possible to get the full picture of the enhancement mechanism by comparing the intensity map with the life-time map. The platform of PDG helps us gain insight into the enhancing mechanism beyond the overall enhancement and provides important information for the design and optimization of the plasmonic nanostructures for surface-enhanced spectroscopy.

**Methods**

**PDG fabrication.** The designed PDGs were fabricated on monocrystalline gold flakes with an atomically flat surface prepared by chemical synthesis.[43] The atomically flat surface of the gold flakes is important because it avoids the noise from unwanted random hot spots due to surface roughness. The Au flakes were then transferred onto ITO glass (ITO thickness = 40 nm) substrate for further patterning using gallium focused ion beam (Helios Nanolab DualBeam G3 UC, FEI, USA). The ion beam current and acceleration voltage were set to 7.7 pA and 3.0 kV, respectively. The fabricated PDGs on gold flakes were further covered by R6G and 4-ATP molecules for SEF/SERS and SECARS, respectively. For SEF/SERS, the R6G doped in a PMMA solution ($10^{-4}$ M, dissolved in chlorobenzene with 0.5% PMMA) was spin-coated onto the surface of the PDG. For SECARS, the fabricated PDG was dipped into a concentric solution of 4-ATP (10 mM, dissolved in methanol) overnight to form a dense monolayer of 4-ATP molecules on the PDG[40] and the unpatterned gold surface. The sample was then rinsed with methanol to wash out the unabsorbed molecules.

**SEF/SERS and SECARS mapping.** For linear SEF/SERS, we performed spectral mapping on the area including the PDG and the unpatterned flat gold surface. A circularly polarized CW laser at 633 nm was focused onto the sample plane of the PDG by an air objective (63×, NA = 0.90, N-Achroplan, Carl Zeiss, Germany). The size of the focal spot is estimated to be about 430 nm. The focal spot was scanned across the sample by a home-build galvo scanner and the scattering spectrum at each excitation position was recorded. Note that circularly polarized illumination was used to eliminate any possible anisotropy to the grating direction. The emitted photons were collected by the same objective and aligned into the entrance slit of a spectrometer (Andor Shamrock 193i, Oxford Instruments, UK) to record the spectrum at each excitation position. A notch filter and a long-pass filter were used to reject the elastic scattering of the excitation laser. The schematics of the optical setup for SEF/SERS and SECARS mapping can be found in the Supporting Information. For the SECARS experiment, the sample was scanned against the fixed laser focal spot by a piezo stage and the emission intensity at the anti-Stokes frequency was recorded with respect to the stage position. The pump and Stokes beams were produced with a tunable picosecond optical parametric oscillator (OPO) (Levante, Berlin, Germany). The wavelength of the pump beam was set at 955 nm (7 ps, 80 MHz) and that of the Stokes beam at 1064 nm (7 ps, 80 MHz). These two input beams were aligned into a linear polarizer followed by a Berek compensator to produce circularly polarized pulses, which were then focused onto the PDG structure with a 50/50 beam splitter and an IR objective (M Plan Apo NIR, 20×, NA = 0.40, Mitutoyo, Japan). A set of two short-pass filters (FES0900, Thorlabs, USA) and a bandpass filter (BP850-40, Thorlabs, USA) was inserted in front of an avalanche photon counter (PDM series photon counting detector, Micro Photon Devices, Italy) to record the intensity of light at the anti-Stokes frequency. Since plasmonic gratings respond best to *p*-polarized light and the PDG is a circular grating, it is important to keep a high degree of circular polarization of the pump and Stokes beams. Deviation from circular polarization to elliptical polarization would lead to distortion of the mapping image, *i.e.*, asymmetry about the middle line of the PDG.[5, 32] The optical setup for SECARS mapping is provided in the Supporting Information. Note that, for SEF/SERS, we have chosen to scan the focal spot against the sample by a galvo scanner in the experiment because it is faster. This is possible for SEF/SERS because the signal intensity is rather high and the distortion of the focal spot due to the laser scanning is not critical for SEF/SERS. Differently, for SECARS, we have chosen to scan the sample against a fixed laser focal spot by a piezo stage. This is a relatively slow method but allows us to maintain the excitation condition. This is of critical importance for SECARS because SECARS requires two beams (pump and Stokes) to be aligned and focused onto the sample with good spatial and temporal overlap, as well as circular polarization. Using a galvo scanner to scan the focal spot can lead to an unwanted variation of the excitation condition

(*e.g.*, variation in the spatial overlap or the spatial phase of the two beams) and thus systematic error in the CARS signal.

ASSOCIATED CONTENT

**Supporting Information**. DFT calculation, experimental setup, video of SERS image evolution, FDTD simulations, and pixel-to-azimuthal angle conversion are included in Supporting Information. This material is available free of charge *via* the Internet at http://pubs.acs.org.

Additional figures (PDF)

AUTHOR INFORMATION

**Corresponding Author**

\* E-mail: jer-shing.huang@leibniz-ipht.de

**Author Contributions**

J.-S.H. conceived the research idea. L.O., F.-C.L., K.-M.S., and J.-S. H. designed and fabricated the sample. L.O., T.M, K.-M.S., W.-L.C. F.-C.L. D.A., and Y.-M.C performed the optical experiment. L.O. and K.-M.S. performed theoretical analysis and FDTD simulations. S.E., M.R., and S.G. performed DFT calculations. M.S., Y.-M.C., S.G., J.P., and J.-S.H. supervised the research. L.O. and J.-S.H. wrote the manuscript. All authors contributed to the interpretation of the experimental results and the revision of the manuscript.

ACKNOWLEDGMENT

Financial support from the Thuringian State Government within its ProExcellence initiative (APC$^{2020}$) and third-party funds from Deutsche Forschungsgemeinschaft (Subprojects A4 and C1 of CRC 1375 NOA, HU2626/3-1), European Research Council (ERC grant QUEM-CHEM, 772676), and the Ministry of Science and Technology of Taiwan (MOST-103-2113-M-007-004-MY3) are gratefully acknowledged.

REFERENCES

1. Biagioni, P.; Huang, J.-S.; Hecht, B., Nanoantennas for Visible and Infrared Radiation. *Rep. Prog. Phys.* **2012,** *75*, 024402.

2. Zhan, C.; Chen, X.-J.; Yi, J.; Li, J.-F.; Wu, D.-Y.; Tian. Z.-Q., From Plasmon-Enhanced Molecular Spectroscopy to Plasmon-Mediated Chemical Reactions. *Nat. Rev. Chem.* **2018,** *2*, 216-230.

3. Clavero, C., Plasmon-Induced Hot-Electron Generation at Nanoparticle/Metal-Oxide Interfaces for Photovoltaic and Photocatalytic Devices. *Nature Photonics* **2014,** *8*, 95-103.

4. Wong, C. L.; Olivo, M., Surface Plasmon Resonance Imaging Sensors: A Review. *Plasmonics* **2014,** *9*, 809-824.

5. Lin, F.-C.; See, K.-M.; Ouyang, L.; Huang, Y.-X.; Chen, Y.-J.; Popp, J.; Huang, J.-S., Designable Spectrometer-Free Index Sensing Using Plasmonic Doppler Gratings. *Anal. Chem.* **2019,** *91*, 9382-9387.

6. Aroca, R. F.; Ross, D. J.; Domingo, C., Surface-Enhanced Infrared Spectroscopy. *Appl. Spectrosc.* **2004,** *58*, 324A-338A.

7. Domke, K. F.; Zhang, D.; Pettinger, B., Toward Raman Fingerprints of Single Dye Molecules at Atomically Smooth Au(111). *J. Am. Chem. Soc.* **2006,** *128*, 14721-14727.

8. Fort, E.; Grésillon, S., Surface Enhanced Fluorescence. *J. Phys. D: Appl. Phys.* **2008,** *41*, 13001-31.

9. Zhang, R.; Zhang, Y.; Dong, Z. C.; Jiang, S.; Zhang, C.; Chen, L. G.; Zhang, L.; Liao, Y.; Aizpurua, J.; Luo, Y.; Yang, J. L.; Hou, J. G., Chemical Mapping of a Single Molecule by Plasmon-Enhanced Raman Scattering. *Nature* **2013,** *498*, 82-86.

10. Voronine, D. V.; Zhang, Z.; Sokolov, A. V.; Scully, M. O., Surface-Enhanced FAST CARS: En Route to Quantum Nano-Biophotonics. *Nanophotonics* **2017,** *7*, 523-548.

11. Langer, J.; Jimenez de Aberasturi, D.; Aizpurua, J.; Alvarez-Puebla, R. A.; Auguié, B.; Baumberg, J. J.; Bazan, G. C.; Bell, S. E. J.; Boisen, A.; Brolo, A. G.; Choo, J.; Cialla-May, D.; Deckert, V.; Fabris, L.; Faulds, K.; García de Abajo, F. J.; Goodacre, R.; Graham, D.; Haes, A. J.; Haynes, C. L.*, et al.*, Present and Future of Surface-Enhanced Raman Scattering. *ACS Nano* **2020,** *14*, 28-117.

12. Chu, Y.; Banaee, M. G.; Crozier, K. B., Double-Resonance Plasmon Substrates for Surface-Enhanced Raman Scattering with Enhancement at Excitation and Stokes Frequencies. *ACS Nano* **2010,** *4*, 2804-2810.

13. Alonso-González, P.; Albella, P,; Schnell, M.; Chen, J.; Huth, F.; Garcia-Etxarri, A.; Casanova, F.; Golmar, F.; Arzubiaga, L.; Hueso, L. E.; Aizpurua, J.; Hillenbrand, R., Resolving the Electromagnetic Mechanism of Surface-Enhanced Light Scattering at Single Hot Spots. *Nat. Commun.* **2012,** *3*, 684.

14. Yampolsky, S.; Fishman, D. A.; Dey, S.; Hulkko, E.; Banik, M.; Potma, E. O.; Apkarian, V. A., Seeing a Single Molecule Vibrate through Time-Resolved Coherent Anti-Stokes Raman Scattering. *Nat. Photonics* **2014,** *8*, 650-656.

15. Zhang, Y.; Zhen, Y.-R.; Neumann, O.; Day, J. K.; Nordlander, P.; Halas, N. J., Coherent Anti-Stokes Raman Scattering with Single-Molecule Sensitivity Using a Plasmonic Fano Resonance. *Nat. Commun.* **2014,** *5*, 4424.

16. Gruenke, N. L.; Cardinal, M. F.; McAnally, M. O.; Frontiera, R. R.; Schatz, G. C.; van Duyne, R. P., Ultrafast and Nonlinear Surface-Enhanced Raman Spectroscopy. *Chem. Soc. Rev.* **2016,** *45*, 2263-2290.

17. Shutov, A. D.; Yi, Z.; Wang, J.; Sinyukov, A. M.; He, Z.; Tang, C.; Chen, J.; Ocola, E. J.; Laane, J.; Sokolov, A. V.; Voronine, D. V.; Scully, M. O., Giant Chemical Surface Enhancement of Coherent Raman Scattering on $MoS_2$. *ACS Photonics* **2018,** *5*, 4960-4968.

18. Evans, C. L.; Xie, X. S., Coherent Anti-Stokes Raman Scattering Microscopy: Chemical Imaging for Biology and Medicine. *Annu. Rev. Anal. Chem.* **2008,** *1*, 883-909.

19. Chen, C. K.; de Castro, A. R. B.; Shen, Y. R.; DeMartini, F., Surface Coherent Anti-Stokes Raman Spectroscopy. *Phys. Rev. Lett.* **1979,** *43*, 946-949.

20. Liang, E. J.; Weippert, A.; Funk, J. -M.; Materny, A.; Kiefer, W., Experimental Observation of Surface-Enhanced Coherent Anti-Stokes Raman Scattering. *Chem. Phys. Lett.* **1994,** *227*, 115-120.

21. Ichimura, T.; Hayazawa, N.; Hashimoto, M.; Inouye, Y.; Kawata, S., Local Enhancement of Coherent Anti-Stokes Raman Scattering by Isolated Gold Nanoparticles. *J. Raman Spectrosc.* **2003,** *34*, 651-654.

22. Koo, T.-W.; Chan, S.; Berlin, A. A., Single-Molecule Detection of Biomolecules by Surface-Enhanced Coherent Anti-Stokes Raman Scattering. *Opt. Lett.* **2005,** *30*, 1024-1026.

23. Genevet, P.; Tetienne, J.-P.; Gatzogiannis, E.; Blanchard, R.; Kats, M. A.; Scully, M. O.; Capasso, F., Large Enhancement of Nonlinear Optical Phenomena by Plasmonic Nanocavity Gratings. *Nano Lett.* **2010,** *10*, 4880-4883.

24. Schlücker, S.; Salehi, M.; Bergner, G.; Schütz, M.; Ströbel, P.; Marx, A.; Petersen, I.; Dietzek, B.; Popp, J., Immuno-Surface-Enhanced Coherent Anti-Stokes Raman Scattering Microscopy: Immunohistochemistry with Target-Specific Metallic Nanoprobes and Nonlinear Raman Microscopy. *Anal. Chem.* **2011,** *83*, 7081-7085.

25. Steuwe, C.; Kaminski, C. F.; Baumberg, J. J.; Mahajan, S., Surface Enhanced Coherent Anti-Stokes Raman Scattering on Nanostructured Gold Surfaces. *Nano Lett.* **2011,** *11*, 5339-5343.

26. Feng, Y.; Gao, M.; Wang, Y.; Yang, Z.; Meng, L., Surface and Coherent Contributions of Plasmon Fields to Ultraviolet Tip-Enhanced Coherent Anti-Stokes Raman Scattering. *Nanotechnology* **2020,** *31*, 395204.

27. Crampton, K. T.; Zeytunyan, A.; Fast, A. S.; Ladani, F. T.; Alfonso-Garcia, A.; Banik, M.; Yampolsky, S.; Fishman, D. A.; Potma, E. O.; Apkarian, V. A., Ultrafast Coherent Raman Scattering at Plasmonic Nanojunctions. *J. Phys. Chem. C* **2016,** *120*, 20943-20953.

28. He, J.; Fan, C.; Ding, P.; Zhu, S.; Liang, E., Near-Field Engineering of Fano Resonances in a Plasmonic Assembly for Maximizing CARS Enhancements. *Sci. Rep.* **2016,** *6*, 20777.

29. Wang, J.; Zhang, J.; Tian, Y.; Fan, C.; Mu, K.; Chen, S.; Ding, P.; Liang, E., Theoretical Investigation of a Multi-Resonance Plasmonic Substrate for Enhanced Coherent Anti-Stokes Raman Scattering. *Opt. Express* **2017,** *25*, 497-507.

30. Fabelinsky, V. I.; Kozlov, D. N.; Orlov, S. N.; Polivanov, Y. N.; Shcherbakov, I. A.; Smirnov, V. V.; Vereschagin, K. A.; Arzumanyan, G. M.; Mamatkulov, K. Z.; Afanasiev, K. N.; Lagarkov, A. N.; Ryzhikov, I. A.; Sarychev, A. K.; Budashov, I. A.; Nechaeva, N. L.; Kurochkin, I. N., Surface-Enhanced Micro-CARS Mapping of a Nanostructured Cerium Dioxide/Aluminum Film Surface with Gold Nanoparticle-Bound Organic Molecules. *J. Raman Spectrosc.* **2018,** *49*, 1145-1154.

31. Zhang, J.; Chen, S.; Wang, J.; Mu, K.; Fan, C.; Liang, E.; Ding, P., An Engineered CARS Substrate with Giant Field Enhancement in Crisscross Dimer Nanostructure. *Sci. Rep.* **2018,** *8*, 740.

32. See, K.-M.; Lin, F.-C.; Huang, J.-S., Design and Characterization of a Plasmonic Doppler Grating for Azimuthal Angle-Resolved Surface Plasmon Resonances. *Nanoscale* **2017,** *9*, 10811-10819.

33. Laux, E.; Genet, C.; Skauli, T.; Ebbesen, T. W., Plasmonic Photon Sorters for Spectral and Polarimetric Imaging. *Nat. Photonics* **2008,** *2*, 161-164.

34. Barnes, W. L.; Dereux, A.; Ebbesen, T. W., Surface Plasmon Subwavelength Optics. *Nature* **2003,** *424*, 824-830.

35. Deng, X.; Braun, G. B.; Liu, S.; Sciortino, P. F.; Koefer, B.; Tombler, T.; Moskovits, M., Single-Order, Subwavelength Resonant Nanograting as A Uniformly Hot Substrate for Surface-Enhanced Raman Spectroscopy. *Nano Lett.* **2010,** *10*, 1780-1786.

36. Jiang, Y.; Wang, H.-Y.; Wang, H.; Gao, B.-R.; Hao, Y.-w.; Jin, Y.; Chen, Q.-D.; Sun, H.-B., Surface Plasmon Enhanced Fluorescence of Dye Molecules on Metal Grating Films. *J. Phys. Chem. C* **2011,** *115*, 12636-12642.

37. Kazemi-Zanjani, N.; Shayegannia, M.; Prinja, R.; Montazeri, A. O.; Mohammadzadeh, A.; Dixon, K.; Zhu, S.; Selvaganapathy, P. R.; Zavodni, A.; Matsuura, N.; Kherani, N. P., Multiwavelength Surface-Enhanced Raman Spectroscopy Using Rainbow Trapping in Width-Graded Plasmonic Gratings. *Adv. Opt. Mater.* **2018,** *6*, 1701136.

38. Andrade, G. F. S.; Min, Q.; Gordon, R.; Brolo, A. G., Surface-Enhanced Resonance Raman Scattering on Gold Concentric Rings: Polarization Dependence and Intensity Fluctuations. *J. Phys. Chem. C* **2012,** *116*, 2672-2676.

39. Kan, Y.; Ding, F.; Zhao, C.; Bozhevolnyi, S. I., Directional Off-Normal Photon Streaming from Hybrid Plasmon-Emitter Coupled Metasurfaces. *ACS Photonics* **2020,** *7*, 1111-1116.

40. Hayes, W. A.; Shannon, C., Electrochemistry of Surface-Confined Mixed Monolayers of 4-Aminothiophenol and Thiophenol on Au. *Langmuir* **1996,** *12*, 3688-3694.

41. Ding, S.-Y.; Yi, J.; Li, J.-F.; Ren, B.; Wu, D.-Y.; Panneerselvam, R.; Tian, Z.-Q., Nanostructure-Based Plasmon-Enhanced Raman Spectroscopy for Surface Analysis of Materials. *Nat. Rev. Mater.* **2016,** *1*, 16021.

42. Semaltianos, N. G., Spin-Coated Pmma Films. *Microelectron. J.* **2007,** *38*, 754-761.

43. Huang, J.-S.; Callegari, V.; Geisler, P.; Brüning, C.; Kern, J.; Prangsma, J. C.; Wu, X.; Feichtner, T.; Ziegler, J.; Weinmann, P.; Kamp, M.; Forchel, A.; Biagioni, P.; Sennhauser, U.; Hecht, B., Atomically Flat Single-Crystalline Gold Nanostructures for Plasmonic Nanocircuitry. *Nat. Commun.* **2010,** *1*, 150.

**TOC Figure**

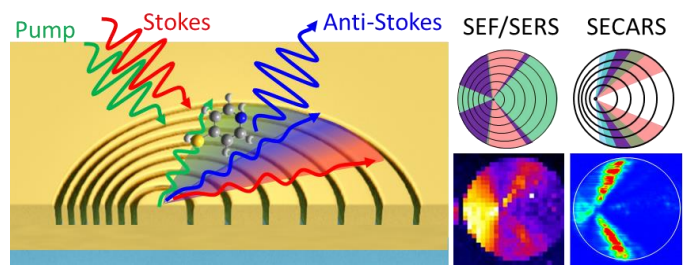

*Supporting Information*

# Spatially Resolving the Enhancement Effect in Surface-Enhanced Coherent Anti-Stokes Raman Scattering by Plasmonic Doppler Gratings

Lei Ouyang[†,‡,§] Tobias Meyer[†,‡], Kel-Meng See[⊥], Wei-Liang Chen[‖], Fan-Cheng Lin[⊥], Denis Akimov[†,‡], Sadaf Ehtesabi[‡], Martin Richter[‡], Michael Schmitt[†,‡], Yu-Ming Chang[‖], Stefanie Gräfe[‡], Jürgen Popp[†,‡], Jer-Shing Huang[†,‡,⊥,#,Δ,*]

[†] Leibniz Institute of Photonic Technology, Albert-Einstein Str. 9, 07745 Jena, Germany
[‡] Institute of Physical Chemistry and Abbe Center of Photonics, Friedrich-Schiller-Universität Jena, Helmholtzweg 4, D-07743 Jena, Germany
[§] School of Chemistry and Chemical Engineering, Huazhong University of Science and Technology, Wuhan 430074, China
[⊥] Department of Chemistry, National Tsing Hua University, 101 Sec. 2, Kuang-Fu Road, Hsinchu 30013, Taiwan
[‖] Center for Condensed Matter Sciences, National Taiwan University, Taipei 10617, Taiwan
[#] Research Center for Applied Sciences, Academia Sinica, 128 Sec. 2, Academia Road, Nankang District, Taipei 11529, Taiwan
[Δ] Department of Electrophysics, National Chiao Tung University, Hsinchu 30010, Taiwan
*Corresponding Author, E-mail: jer-shing.huang@leibniz-ipht.de

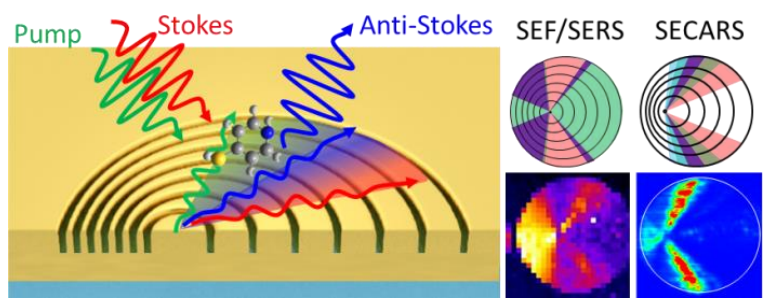

**Content:**

I. DFT calculation (Figure S1)
II. Experimental setup (Figure S2)
III. Video of SERS image evolution (video for PDG-SERS.mp4)
IV. FDTD simulations (Figures S3 and S4)
V. Pixel-to-azimuthal angle conversion (Figure S5)

# I. DFT calculation

In this section, we show by Density Functional Theory (DFT) calculations that the characteristic Raman peak used in this work (1070 cm$^{-1}$) is significantly enhanced by the binding of the 4-Aminothiophenol (4-ATP) molecule to the Au surface. All geometry optimizations, including the cases of isolated 4-ATP (Figure S1a) and 4-ATP attached to the Au surface in 3 different possible orientations (Figure. S1b-S1d), were done using DFT as implemented in GPAW code[1,2] with ASE[3] interface. For this purpose, the optB88-vdW functional[4] is employed in a real-space grid.

The Au slab is represented by a 4×4×3 fcc(111) surface, resulting in 3 layers of 16 Au atoms with two-dimensional periodic boundary conditions (x and y directions). 4-ATP molecules bind strongly to the Au surface through the Au-S bond. To reduce the computational cost of the geometry optimizations, the second and third layer of the Au slab were fixed, and only the geometry of the molecule and Au atoms in the first layer were optimized.

We simulated 3 representative orientations of the 4-ATP molecule with respect to the surface. The first orientation is shown in the inset of Figure S1b, where the molecule binds to the surface through the S atom, and the plane of the phenyl ring is perpendicular to the Au surface. We call this configuration "structure 1". The second orientation is shown in the inset of Figure S1c, where the molecule binds to the surface through the S and H atom. The plane of the phenyl ring is still perpendicular to the Au surface. This configuration is called "structure 2". The third orientation is a "flat-lying" configuration as shown in the insets of Figure S1d. In this configuration, the phenyl ring is parallel to the gold surface. Therefore, the molecule binds to the surface through the interaction of the gold surface with the S atom and with the π orbitals of the phenyl ring. We call this configuration "structure 3".

To obtain the Raman spectra, the vibrational frequencies were calculated using DFT employing the CAM-B3LYP[5] functional and def2-tzvp basis set using Gaussian 09.[6] The frequencies were scaled by a factor of 0.95 to correct for the lack of anharmonicity and the approximate treatment of electron correlation.[7] All vibrational frequency calculations were performed including D3 dispersion correction with Becke-Johnson damping.[8] Figure S1a shows the calculated Raman spectrum of an isolated 4-ATP molecule. Figures S1b to S1d show the calculated spectra of the corresponding configuration shown in their insets. By comparing Figure S1a with Figure. S1b and S1c, it is clear that the characteristic Raman peak at 1070 cm$^{-1}$, which stems from the vibration of the phenyl ring is significantly enhanced by binding of the 4-ATP molecule to the Au surface. By comparing Figure. S1b and S1c with S1d, it is further obvious that 4-ATP molecules with a perpendicular orientation with respect to the Au surface (structure 1 and structure 2) show an approximately three orders of magnitude stronger Raman signal at 1070 cm$^{-1}$ than that of a flat-lying molecule (structure 3). Figure S1e compares the Raman spectra shown in Figure. S1a to S1d. Apparently, attaching the 4-ATP to the gold surface greatly enhances the 1070 cm$^{-1}$ Raman peak, standing out from the background. Therefore, this peak has been chosen as a targeted peak in our SECARS experiment.

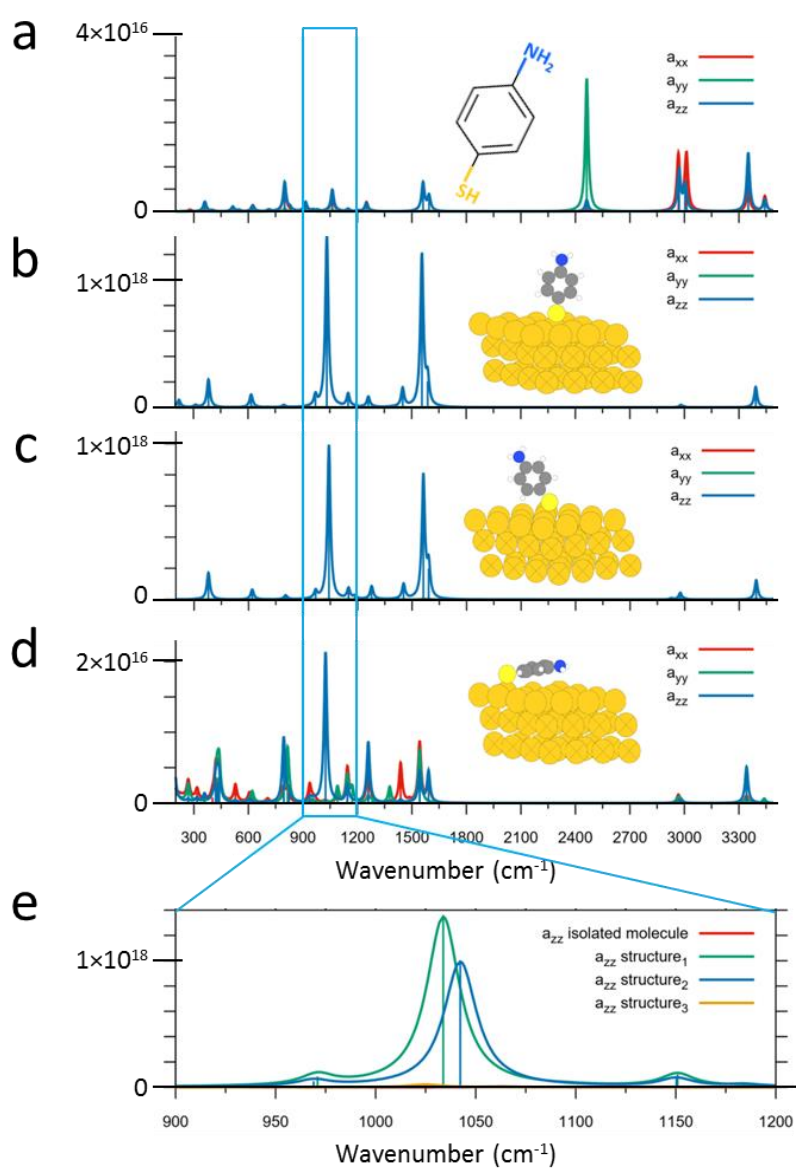

**Figure S1.** (a) Calculated Raman spectrum of an isolated 4-ATP molecule. (b) Calculated Raman spectrum of "structure 1" configuration in the inset. (c) Raman spectrum of "structure 2" configuration in the inset. (d) Raman spectrum of "structure 3" configuration in the inset. Legends in (a) to (d) indicate the different components of the polarizability tensor α, where the coordinate system is such defined that the z-direction points normal to the gold surface. (e) The 1070 cm$^{-1}$ peak of isolated 4-ATP and 4-ATP attached to the Au surface in different orientations.

## II. Experimental setup

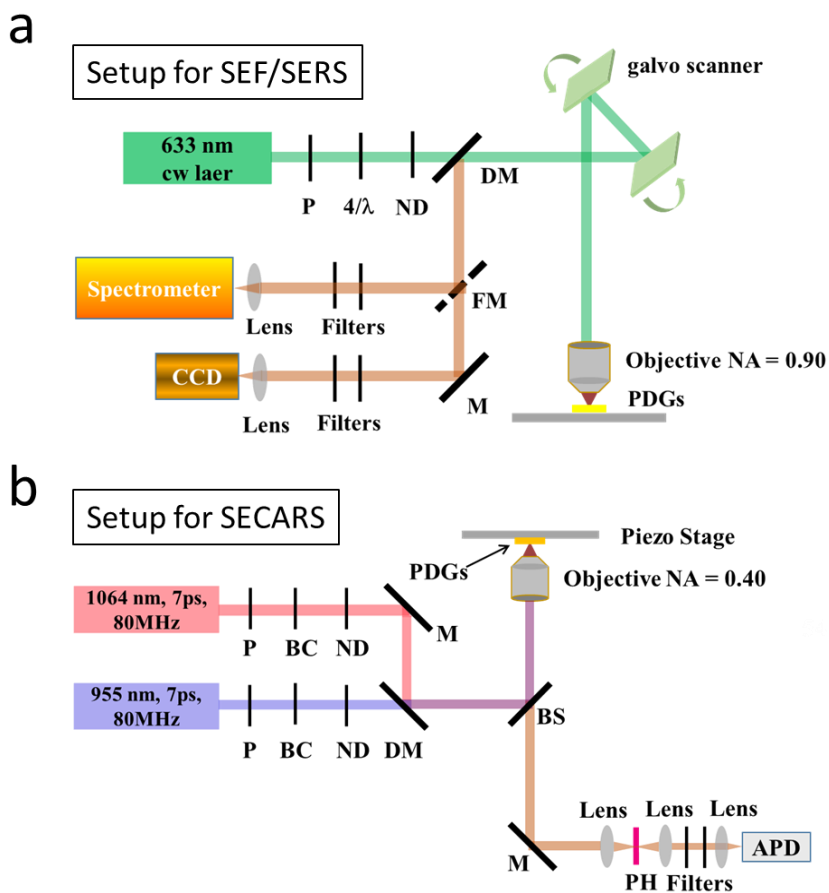

**Figure S2.** (a) The setup for SERS mapping. P: polarizer, 4/λ: quarter wave-plate, ND: neutral density filter, M: mirror, DM: dichroic mirror, FM: flip mirror. (b) The setup for SECARS intensity mapping. P: polarizer, BC: Berek compensator, ND: neutral density filter, M: mirror, DM: dichroic mirror, BS: 50/50 beam splitter, FM: flip mirror, PH: pinhole.

## III. Video of SERS image evolution

"video for PDG-SERS.mp4"

The movie shows the evolution of the SEF/SERS map as the Stokes shift is scanned from 400 to 2650 cm$^{-1}$, corresponding to the spectrum and images shown in Figures 3d and 3e, respectively, in the main text.

## IV. FDTD simulations

### a. Enhancement in the input beams (pump and Stokes)

Planewave source was used to illuminate onto the structure with incident angle increasing from 0 to 23.6°, corresponding to the allowed angle provided by a NA 0.4 objective. The grating period was scanned from 300 to 1100 nm, three point-monitors were placed at three representative positions, which are 5 nm above the structure surface, as shown in Figure S3a. The electric field ($E^2$) intensity recorded by the point monitors was then divided by $E_0^2$ to obtain the enhancement factor. The $E_0^2$ is the electric field recorded by a point monitor placed 5 nm above a flat Au surface. The enhancement factors at the three locations were then used to calculate the averaged enhancement factor for the molecular monolayer covering the surface of the grating. The averaged enhancement factor is calculated area using the formula $\left(\frac{E^2}{E_0^2}\right)_{\text{average}} = \frac{\frac{1}{2}\left(\frac{E_1^2}{E_0^2}+\frac{E_3^2}{E_0^2}\right)*groove\ area + \frac{1}{2}\left(\frac{E_1^2}{E_0^2}+\frac{E_2^2}{E_0^2}\right)*flat\ area}{groove\ area + flat\ area}$.

Taking the dimensions of the structure estimated according to the SEM image (bottom panel of Figure S3a), the formula becomes $\left(\frac{E^2}{E_0^2}\right)_{\text{average}} = \frac{\frac{1}{2}\left(\frac{E_1^2}{E_0^2}+\frac{E_3^2}{E_0^2}\right)*2*\sqrt{(35^2+140^2)} + \frac{1}{2}\left(\frac{E_1^2}{E_0^2}+\frac{E_2^2}{E_0^2}\right)*(P-70)}{2*\sqrt{(35^2+140^2)}+(P-70)}$.

The averaged enhancement was replotted *versus* the azimuthal angle and incident angle for the pump and Stokes beams in Figure. S3b and S3c, respectively. The enhancement factor at a specific azimuthal angle is obtained by averaging the enhancement factor of all possible incident angles. This results in the angle profiles for pump and Stokes shown in Figure 4(c) in the main text.

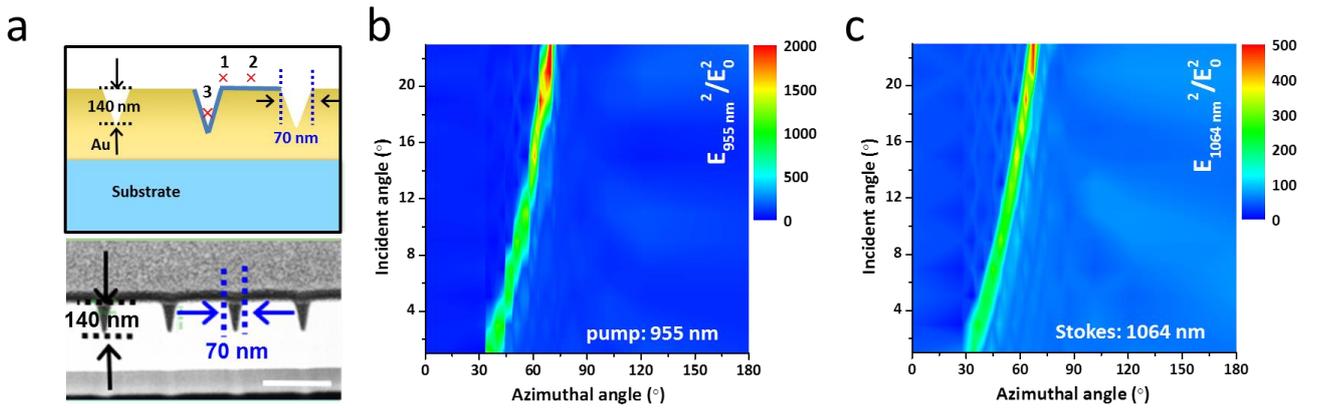

**Figure S3.** (a) Top panel: the grating structure used in the FDTD simulations. The three point-monitors were placed 5 nm above the gold surface at the positions marked by red crosses with a number. Bottom panel: the cross-sectional TEM image of the real structure. The gap width and depth are estimated to be around 70 nm and 140 nm, respectively. (b) and (c): The averaged enhancement factor for the pump and Stokes beam at 955 nm and 1064 nm, respectively, on the 2D plane of the incident and azimuthal angle. Scale bar = 500 nm.

**b. Enhancement in the output beam (anti-Stokes)**

A dipole source was placed 5 nm above the structure at three different locations (Figure S4a). The polarization of the dipole source is in-plane along the x-axis, *i.e.*, perpendicular to the grating grooves. A line monitor was placed 20 nm above the upper surface of the structure. The grating period was scanned from 300 to 1100 nm. The far-field projection of the electric field intensity recorded at 867 nm (anti-Stokes) by the line monitor allows for the calculation of the emission power as a function of the emitting angle. The azimuthal and emission angle-dependent electric field intensity with a single dipole source placed at three representative positions is shown in Figure S4b. The collectible emitted electric field intensity ($E^2$) at a specific azimuthal angle (*i.e.* a specific grating periodicity) is obtained by summing up the electric field intensity for all allowed emission angles defined by the numerical aperture of the objective (NA = 0.4), *i.e.*, the angle between ± 23.6˚. Enhancement $E^2/E_0^2$ was calculated by dividing the $E^2$ with $E_0^2$, where $E_0^2$ is the emitted power obtained with a dipole source on a flat gold surface. The area-averaged enhancement factor was calculated in a way similar to the calculation for the enhancement in the input beams. The area-averaged enhancement plotted *versus* the azimuthal and emission angle is shown in Figure S4b. The enhancement factor at a specific azimuthal angle is obtained by summing up the enhancement factor for all allowed emission angles within the numerical aperture of the objective (NA = 0.4), *i.e.*, all angles between ± 23.6˚. This produces the enhancement angle profiles for the anti-Stokes beam shown in Figure 4(c) in the main text.

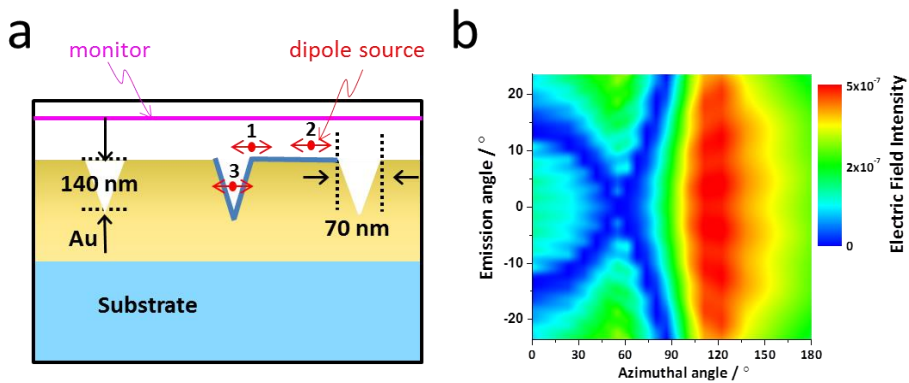

**Figure S4.** (a) The grating structure for simulating the output enhancement of the anti-Stokes beam (867 nm). Three double arrows indicate the three representative positions and orientation of a single dipole source on the grating. A line monitor (magenta line) was placed 20 nm above the upper surface of the structure to record the emission power from the dipole source. (b) The azimuthal angle and emission angle-dependent electric field intensity distribution for anti-Stokes output at 867 nm with the dipole placed at point 2.

## V. Pixel-to-azimuthal angle conversion

To directly compare the angle distribution of the experimental SECARS intensity with the simulated enhancement, the coordinate of each intensity pixel in the intensity map (Figure 4a in the main text) was transformed into the azimuthal angle ($\varphi$) with respect to the center of the PDG, *i.e.*, the position of the smallest ring with a zero radius ($x_0$, $y_0$) using $\varphi = \arctan\frac{|y_n - y_0|}{|x_n - x_0|}$. Figure S5 illustrates the pixel-to-azimuthal angle conversion.

The calculated azimuthal angle for each pixel was rounded to the nearest integer of one degree. Intensities of all pixels within the bandwidth of one degree were summed up and divided by the pixel number to obtain the averaged intensity, as shown by the red trace in Figure 4d in the main text.

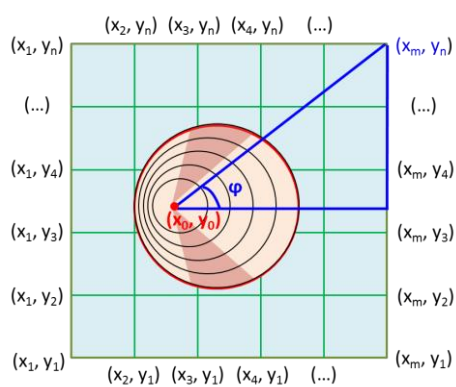

**Figure S5.** Schematic illustration of the pixel-to-azimuthal angle conversion. For the clarity of illustration, the size of pixels has been enlarged.

**REFERENCES**

1. Mortensen, J. J.; Hansen, L. B.; Jacobsen, K. W., Real-Space Grid Implementation of the Projector Augmented Wave Method. *Phys. Rev. B* **2005,** 71, 035109.
2. Enkovaara, J.; Rostgaard, C.; Mortensen, J. J.; Chen, J.; Dulak, M.; Ferrighi, L.; Gavnholt, J.; Glinsvad, C.; Haikola, V.; Hansen, H. A.; Kritstoffersen, H. H.; Kuisma, M.; Larsen, A. H.; Lehtovaara, L.; Ljungberg, M.; Lopez-Acevedo, O.; Moses, P. G.; Ojanen, J.; Olsen, T.; Petzold, V., *et al*., Electronic Structure Calculations with GPAW: A Real-Space Implementation of the Projector Augmented-Wave Method. *J. Phys.: Condens. Matter* **2010,** 22, 253202.
3. Larsen, A. H.; Mortensen, J. J.; Blomqvist, J.; Castelli, I. E.; Christensen, R.; Dulak, M.; Friis, J.; Groves, M. N.; Hammer, B.; Hargus, C.; Hermes, E. D.; Jennings, P. C.; Jensen, P. B.; Kermode, J.; Kitchin, J. R.; Kolsbjerg, E. L.; Kubal, J.; Kaasbjerg, K.; Lysgaard, S.; Moronsson, J. B., *et al.*, The Atomic Simulation Environment - A Python Library for Working with Atoms. *J. Phys.: Condens. Matter* **2017**, 29, 273002.
4. Klimeš, J.; Bowler, D. R.; Michaelides, A, Chemical Accuracy for the van der Waals Density Functional. *J. Phys.: Conden. Matter* **2010**, 22, 022201.
5. Yanai, T.; Tew, D. P.; Handy, N. C., A New Hybrid Exchange–Correlation Functional Using the Coulombattenuating Method (CAM-B3LYP). *Chem. Phys. Lett.* **2004**, 393, 51–57.
6. Frisch, M. J.; Trucks, G. W.; Schlegel, H. B., Scuseria, G. E.; Robb, M. A.; Cheeseman, J. R.; Scalmani, G.; Barone, V.; Mennucci, B.; Petersson, G. A.; Nakatsuji, H.; Caricato, M.; Li, X.; Hratchian, H. P.; Izmaylov, A. F.; Bloino, J.; Zheng, G.; Sonnenberg, J. L.; Hada, M.; Ehara, M., *et al.,* Gaussian 09, Revision A.1, Gaussian, Inc., Wallingford, CT, **2009**.
7. Merrick, J. P.; Moran, D.; Radom, L., An Evaluation of Harmonic Vibrational Frequency Scale Factors. *J. Phys. Chem.* **2007**, 111, 11683–11700.
8. Grimme, S.; Ehrlich, S.; Goerigk, L., Effect of the Damping Function in Dispersion Corrected Density Functional Theory. *J. Comput. Chem*. **2011**, 32, 1456–1465.